\documentstyle[prb,aps,epsf,preprint]{revtex}
\begin{document}
\draft

\title{Work functions of pristine and alkali-metal intercalated carbon nanotubes and bundles}

\author{Jijun Zhao $^a$$^*$, Jie Han $^b$, Jian Ping Lu$^a$ $^{\dagger}$}

\address{$^a$ Department of Physics and Astronomy, University of North
Carolina at Chapel Hill, Chapel Hill, NC 27599 \\
$^b$ NASA Ames Research Center, Mail Stop T27A-1, Moffett Field, CA 94035}

\date{\today} 
\maketitle 

\begin{abstract}

The work functions (WF) of single-walled carbon nanotubes and bundles are studied using first principles methods. For individual metallic tubes, the WF is independent of the chirality and increase slightly with tube diameter. For semiconducting tubes, the WF (as defined by the HOMO energy) decreases rapidly. The WF of nanotube bundles ($\sim$ 5 eV) shows no clear dependence on the tube size and chirality, slightly higher than individual tubes. Calculations on finite tubes show no substantial difference in the tube end and the side wall. Upon alkali-metal intercalation, the WF decreases dramatically and the electronic states near the Fermi level are significantly modified. The metallic and semiconducting nanotubes bundles become indistinguishable.
\end{abstract}


The work function is one of the critical quantities in understanding the field emission properties of carbon nanotubes \cite{1,2,3,4,5}. Although the work function can be estimated from the field-emission spectra based on Fowler-Nordheim model \cite{3,4}, the results are not reliable due to the uncertainty of the local geometry of nanotubes \cite{4}. Other experimental measurements on work functions of both single-walled nanotubes (SWNTs) and multi-walled nanotubes (MWNTs) included ultraviolet photoemission spectroscopy (UPS) \cite{6,7,8,9,10,11} and transmission electron microscopy \cite{12}.  It was found that the work functions of MWNTs are about 0.1-0.2 eV lower than that of the graphite \cite{6,7,10,12}, while the SWNT bundles have slightly higher work functions \cite{8,9}. Upon Cs intercalation the work functions of carbon nanotubes are reduced dramatically \cite{9,11}, which leads to a significant enhancement in field emission \cite{5}. Up to now, theoretical works were only limited to a few nanotubes with finite-lengths \cite{13,14,15}. In this letter, we report results of first principles calculations on the work functions of individual carbon nanotubes and nanotube bundles. The effect of tube diameter and chirality, and the difference between the tube end and the side wall are investigated. The alkali-metal (K, Rb, Cs) intercalated carbon nanotube bundles are also discussed. 

The work function of a bulk metal is related to its Fermi energy $E_F$ by $WF=\phi-E_F$, where $\phi$ is the electrostatic potential caused by ``spilling out'' of electron density at the metal surface \cite{16,17,18}. For those metals with low electron density such as K, Rb, Cs, it is known that $\phi$ is much smaller than $E_F$ \cite{16,17}. For carbon nanotubes, the conduction electron density is much smaller than that of K or Cs \cite{19}. In this work, we approximate the WF by the Fermi energy $E_F$ \cite{13,14}.
To determine the Fermi level of carbon nanotubes with respect to the vacuum level, we perform all electron LCAO calculations based on the density functional DMol package \cite{20}. The density functional is treated by the generalized gradient approximation (GGA) \cite{21} with the exchange-correlation potential parameterized by Wang and Perdew (PW91) \cite{22}. For the infinite nanotubes, one-dimensional (1-D) periodic boundary condition is applied along the tube axis. The Brillouin zone is sampled by large sets of Monkhorst-Pack {\bf k} meshes \cite{23} (Along the tube axis, 40 {\bf k} points are used for standard calculations and 160 points for accurate electronic density of states). Benchmark calculations are carried out on several alkali metals solids and the graphene sheet. The ionization potential of a C$_{60}$ cluster is also calculated from the total energy difference between the neutral and the charged cluster. As shown in Table I, the overall agreement between theory and experiments \cite{8,9,24,25} for different systems is reasonable. 
 
\begin{table}
Table I. Work function of bulk alkali metals (K, Rb, Cs) \cite{24}, graphene sheet \cite{8,9}, nanotube rope (see Table II and discussions in text) \cite{8,9}, and the ionization potentials of C$_{60}$ cluster \cite{25}. GGA denotes present GGA calculations, Exp. are the experimental WF values \cite{8,9,24,25}. All units are in eV.
\begin{center}
\begin{tabular}{cccccccc} 
            &   K     &  Rb     &   Cs   &  graphene & nanorope  & C$_{60}$      \\ \hline
WF (GGA)    &  2.69   &  2.42   &  2.31  &  4.8       &  $\sim$ 5.0  &   7.87          \\
WF (Exp.)   &  2.30   &  2.16   &  2.14  & 4.8 \cite{8}, 4.6 \cite{9} & 5.05 \cite{8}, 4.8 \cite{9} & 7.61\\  
\end{tabular}
\end{center}
\end{table}

We first address the work functions of the individual metallic SWNTs of infinite length, with diameter ranging from 5.7\AA$~$ to 16.3 \AA$~$. Both the armchair ($n$,$n$) and the zigzag ($n$,0) ($n=3m$) chirality are considered. The calculated WFs range from 4.63 to 4.77 eV, given in Fig.1 and Table II. Careful examinations show that the WF of the metallic tubes increases slightly with 1/D (Fig.1). Extrapolation towards the D$\rightarrow \infty$ limit gives a WF$_\infty$ of 4.83 eV, close to the calculated value for the graphene sheet (4.8 eV). It is worthy to note that the WFs of both the armchair and the zigzag metallic tubes fit the same linear dependence, indicating that the WFs of metallic nanotubes are independent of chirality.
\begin{figure}
\vspace{1.3in}
\centerline{
\epsfxsize=5.0in \epsfbox{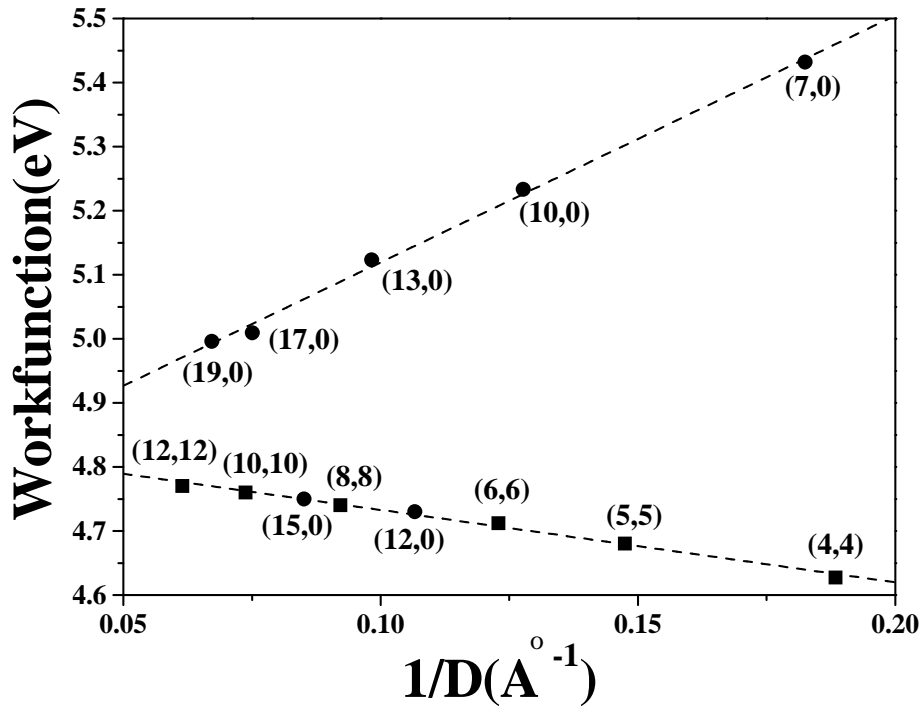}
}
\vspace{-0.75in}
\caption{Work functions (eV) of individual metallic and semiconducting SWNTs vs. the inverse tube diameter 1/D (squares:  armchair SWNTs; dots: zigzag SWNTs). Linear extrapolation towards the D$\rightarrow \infty$ limit yields WF$_\infty$ of 4.84 eV for metallic tubes and 4.73 for semiconducting ones, close to the calculated value (4.8 eV) for graphene sheet.}
\end{figure}

For the purpose of comparison, we define the WF of semiconducting tube as the highest occupied molecular orbital (HOMO) energy. The WFs of the semiconducting tube are substantially higher than the metallic ones (Fig.1). The calculated WF decreases linearly with 1/D and approaches an extrapolation limit 4.73 eV at D$\rightarrow \infty$. The strong diameter dependence come from the well-known gap decrease with tube diameter in semiconducting tubes \cite{26}. There is no direct experiment measurement on the work functions of individual SWNTs. But recent experiments on the MWNT tips suggest that the WFs of semiconducting tubes ($\sim 5.6$ eV) are high than those of metallic ones ($\sim 4.6-4.8$ eV) \cite{12}.

To explore the difference in work function between the side wall and the tube end, we carry out model calculations on several finite (5,5) tubes with length up to 21.8 \AA. As shown in Fig.2, one end of the nanotube is capped by a half-C$_{60}$ while  the other end is H-terminated. The calculated HOMO energy decreases with the tube length $l$ and approaches the infinite limit rapidly (4.87 eV, 4.63 eV, 4.55 eV  for $l=3.8 \AA, 11.7 \AA, 21.8 \AA$ respectively). In Fig.3, we plot the HOMO wave function distributions for the longest tube studied (160 carbon atoms on side wall plus 30 carbon atom on the end). The HOMO are clearly delocalized over the whole tube. The wave function at the tube end is comparable to that on the side wall. Similar distributions can be found for the molecular orbitals just below the HOMO level. Therefore, we expect that the work function at the tube end should not be much different from that of the sidewall. For the finite size nanotubes investigated, we find interesting oscillations in both the HOMO energy and the wave function distribution with the tube length. The oscillation period is three times that of the unit cells length along the tube axis, corresponding to the Fermi wavelength of armchair tubes \cite{27}. This oscillation behavior implies that care should be taken when extrapolating the finite size results to the infinite limit \cite{13,14,15}. Our detailed investigations on the electronic properties of finite nanotubes with different lengths and ends will be presented elsewhere. 

\begin{figure}
\centerline{
\epsfxsize=5.0in \epsfbox{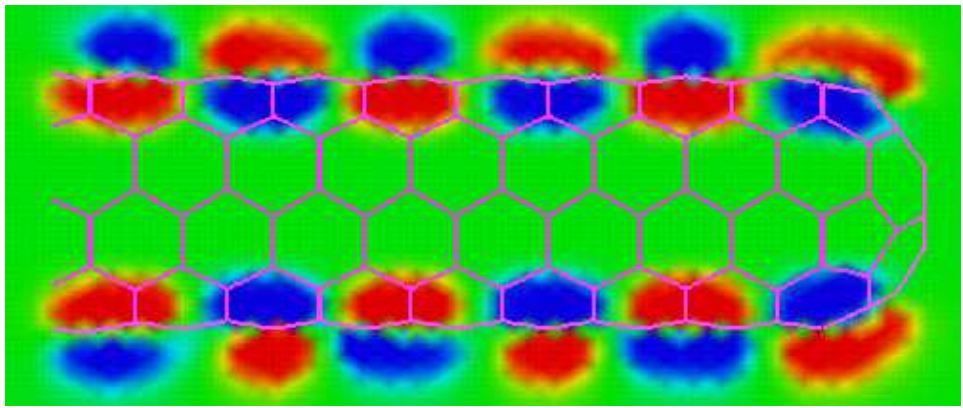}
}
\vspace{0.15in}
\caption{Distribution of HOMO wave function (slice through the middle of nanotube) in a finite length (21.8 \AA$~$) tube with one closed end. The red (blue) color denotes positive (negative) value of the wave function. The delocalization of HOMO is clearly shown. The oscillations of wave function along the tube axis with a period of 3 unit cells correspond to the Fermi wavelength of armchair tubes.}
\end{figure}
\begin{table}

Table II. Work function (eV) of individual metallic tubes and ropes (tube bundle) with various diameters and chirality (m,n) obtained from GGA calculations.
\begin{center}
\begin{tabular}{cccccccc} 
(m,n)   & (5,5) & (6,6) & (12,0) &  (8,8) & (15,0) & (10,10) & (12,12) \\ \hline
tube    & 4.68  & 4.71  &  4.73  &  4.74  & 4.73   &  4.76   &  4.77   \\
rope    & 5.08  & 5.07  &  5.05  &  5.00  & 4.98   &  5.01   &  4.94   \\ 
\end{tabular}
\end{center}
\end{table}

We now discuss the work functions of the nanotube bundles. The bundles are modeled by two dimensional hexagonal lattices of uniform SWNTs \cite{28}. For all the metallic tube bundles studied, the calculated work functions are around 5 eV (see Table II), slightly higher than individual metallic nanotubes and the graphene. The increase of WFs in tube bundles can be understood by the tube-tube interaction \cite{29}. From the experiments based on UPS, the work functions of SWNT bundles are determined to be 4.8 eV \cite{8} or 5.05 eV \cite{9}. Our theoretical results agree well with the available data on nanotube bundles. 

In addition to pristine materials, the electronic properties of carbon nanotubes can be efficiently controlled by alkali-metal intercalations. Recently, nanoelectronics devices have been constructed on the basis of alkali-metal doped carbon nanotubes \cite{30}. Thus, it is important to investigate the effect of alkali-metal intercalations. We have carried out systematical calculations on the effect of intercalations in both metallic (10,10) and semiconducting (17,0) tubes bundles. Intercalation density up to K$_{0.1}$C (Rb, Cs) (close to the saturation density in nanotube bundles \cite{31} and the graphite \cite{32}) are studied. Both the inside of SWNTs and the interstitial sites are explored. The initial configurations of alkali-metal atoms are chosen to maximize the ion-ion distance \cite{28}. Structural relaxations find no significant change from the initial configurations. For a given intercalation concentration, the WF is insensitive to the detailed configurations of the intercalated atoms. The lattice constants of two-dimensional hexagonal lattices are expanded. For example, we find that the intercalations of K atoms into the interstial sites of (10,10) tube bundle can induce about $\sim$2\AA$~$ lattice expansion, which is comparable to expansion of 1.95 \AA$~$ in K intercalated graphite \cite{32} and 1.85 \AA$~$ in the HNO$_3$ intercalated SWNT bundles \cite{33}.
In Fig.3, we present the calculated work functions of intercalated tube bundles as functions of the metal concentration for various metals. In general, the work function dramatically decreases after small amount of alkali-metal intercalations. The reduction becomes much slower at higher intercalation density. Furthermore, there is almost no difference between the WFs of intercalated (10,10) and (17,0) tube bundles. Thus, the metallic and semiconducting tube bundles become indistinguishable. The WF for Cs-doped nanotube bundles is slightly lower than that of the Rb- or K- doped systems. This trend from K to Cs is consistent with the WFs for bulk metal that ($WF(K)>WF(Rb)>WF(Cs)$, see Table I).

\begin{figure}
\vspace{1.5in}
\centerline{
\epsfxsize=5.0in \epsfbox{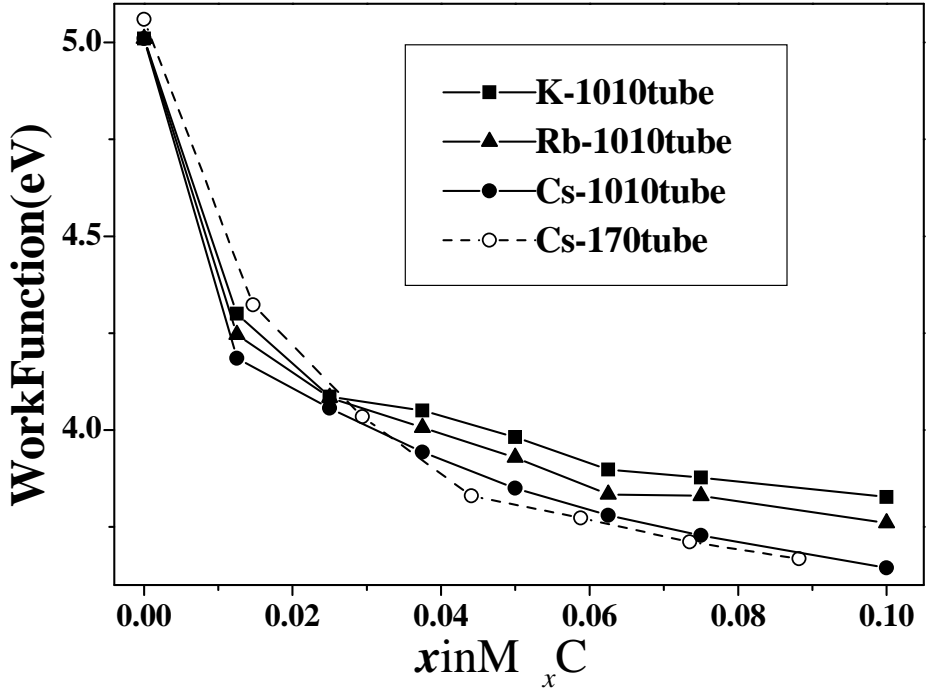}
}
\vspace{-0.75in}
\caption{Work functions (eV) of alkali-metal intercalated carbon nanotube bundles vs. the intercalation density ($x$ in M$_x$C, M denotes metal). Initially, work function decreases dramatically with intercalation density. The reduction of WF becomes much slower at higher density.}
\end{figure}

Intuitively, the reduction of WFs can be understood by the charge transfer from metal to carbon nanotube, which shifts the E$_F$ of conduction band towards the vacuum level. Experimentally, the charge transfer from metal atoms to the nanotube have been confirmed by the resistance \cite{34,35} and Raman spectrum \cite{36}. However, our present results show that interaction between K (or Rb, Cs) and nanotube cannot be simply described by a rigid band model with charge transfer. Shown in Fig.4 is the electronic density states for pristine and K-intercalated (10,10) SWNT bundles. Similar to the case of Li intercalation \cite{28}, the valence bands are almost not affected by K intercalations. In contrast, the conduction bands are significantly modified by potassium-carbon interaction. New peaks associated with alkali-metal atoms are found in the conduction bands. The density of states near the Fermi level is significant enhanced by the contributions of alkali metals. Our thoeretical results are supported by recent experiments on the optical properties of the K or Cs intercalated nanotube bundles \cite{37}. In addition, we have also calculated the density of states for intercalated semiconducting bundles. Similar to the case of work function, the metallic and semiconducting nanotubes bundles become indistinguishable. This is consistent with recent NMR measurements on K-intercalated SWNT bundles \cite{38}.
\begin{figure}
\vspace{1.3in}
\centerline{
\epsfxsize=5.0in \epsfbox{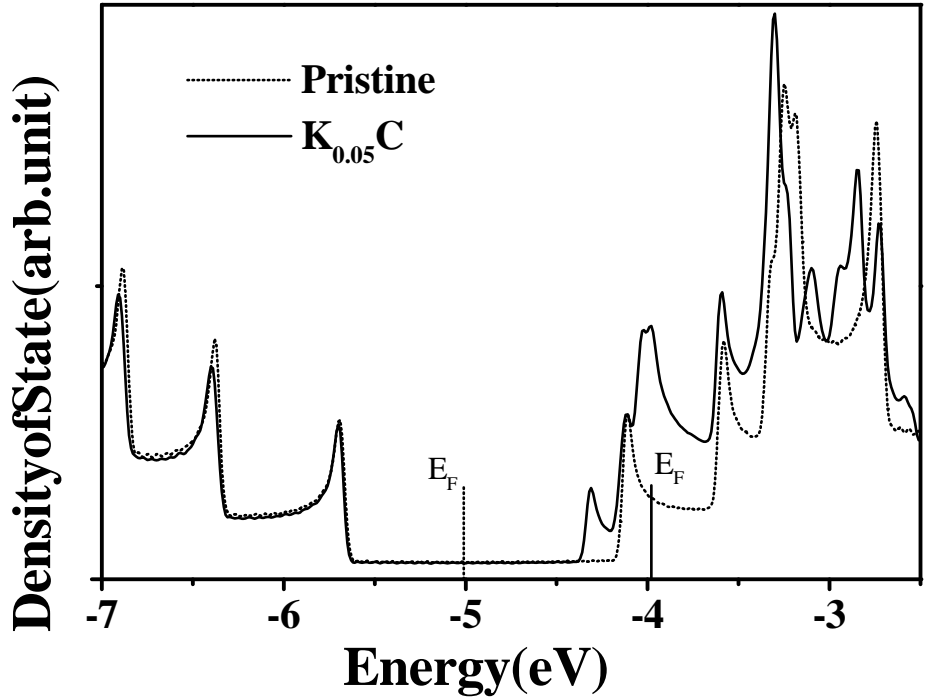}
}
\vspace{-0.75in}
\caption{Electronic density of states of pristine (dotted line) and K-intercalated (solid line) (10,10) SWNT bundles (K$_{0.05}$C). The valence bands of nanotube are almost not affected by K intercalations, while the conduction bands are significantly modified by the potassium-carbon interactions.}
\end{figure}

In summary, we have performed first principles calculations on the work functions of pristine and intercalated SWNT nanotube and bundles. The WFs of metallic nanotubes are weakly depend on the tube size and comparable to the graphene sheet. The work functions of all the metallic tube bundles are around 5 eV, slightly higher than those of individual tubes. Finite size tube calculations reveal the delocalized HOMO state, implying no substantial difference between the WF of the tube end and the side wall. Upon alkali-metal intercalation, the WFs of tube bundles decrease dramatically and the electronic states near Fermi level are significantly modified. The metallic and semiconducting nanotubes bundles become indistinguishable.
This work is supported by the U.S. Army Research Office Grant No. DAAG55-98-1-0298, the Office of Naval Research Grant No. N00014-98-1-0597 and NASA Ames Research Center. The authors thank Prof. O.Zhou, Prof. Y.Wu for helpful discussions. We acknowledge computational support from the North Carolina Supercomputer Center.

\ \\
$^*$E-mail address: zhaoj@physics.unc.edu \\
$^{\dagger}$E-mail address: jpl@physics.unc.edu


\begin{references}

\bibitem{1} R.H.Fowler and L.Nordheim, Proc.R.Soc.London, Ser.A {\bf 119}, 683(1928).

\bibitem{2} W.A.De Heer, A.Chatelain, D.Ugarte, Science{\bf 270}, 1179(1995); W.Zhu, C.Bower, O.Zhou, G.Kochanski, S.Jin, Appl.Phys.Lett.{\bf 75}, 873(1999).
\bibitem{3} M.Tian, L.Chen, F.Li, R.Wang, Z.Mao, Y.Zhang, J.Appl.Phys.{\bf 82}, 3164(1997); O.Groning, O.M.Kuttel, Ch.Emmenegger, P.Groning, L.Schlapbach, J.Vac.Sci.Technol.B{\bf 18}, 665(2000).

\bibitem{4} P.G.Collins, A.Zettl, Phys.Rev.B{\bf 55}, 9391(1997).
\bibitem{5} A.Wadhawan, R.E.Stallcup, J.M.Perez, Appl.Phys.Lett.{\bf 78}, 108(2001).

\bibitem{6} H.Ago, T.Kugler, F.Cacialli, W.R.Salaneck, M.S.P.Shaffer, A.H.Windle, R.H.Friend, J.Phys.Chem.B{\bf 103}, 8116(1999).
\bibitem{7} P.Chen, X.Wu, X.Sun, J.Lim, W.Ji, K.L.Tan, Phys.Rev.Lett.{\bf 82}, 2548(1999).
\bibitem{8} M.Shiraishi, M.Ata, Carbon {\bf 39}, 1913(2001). 
\bibitem{9} S.Suzuki, C.Bower, Y.Matanabe, O.Zhou, Appl.Phys.Lett.{\bf 76}, 4007(2000).
\bibitem{10} S.Suzuki, Y.Matanabe, T.Kiyokura, K.G.Nath, T.Ogino, S.Heun, W.Zhu, C.Bower, O.Zhou, Phys.Rev.B{\bf 63}, 245418(2001).
\bibitem{11} S.Suzuki, Y.Matanabe, T.Kiyokura, K.G.Nath, T.Ogino, S.Heun, W.Zhu, C.Bower, O.Zhou, unpublished.
\bibitem{12} R.Gao, Z.Pan, Z.L.Wang, Appl.Phys.Lett.{\bf 78}, 1757(2001).
\bibitem{13} G.Zhou, W.Duan, B.Gu, Appl.Phys.Lett.{\bf 79}, 836(2001).
\bibitem{14} G.Zhou, W.Duan, B.Gu, Phys.Rev.Lett.{\bf 87}, 095504(2001).
\bibitem{15} A.Maiti, J.Andelm, N.Tanpipat, P.von Allmen, Phys.Rev.Lett.{\bf 87}, 155502(2001).
\bibitem{16} N.D.Lang, W.Kohn, Phys.Rev.B{\bf 3}, 1215(1971).
\bibitem{17} H.L.Skriver, N.M.Bosengaard, Phys.Rev.B{\bf 46}, 7157(1992).
\bibitem{18} N.W.Ashcroft and N.D.Mermin, {\em Solid State Physics}, (Saunders College, Philadelphia, 1976).

\bibitem{19} In the (10,10) tube bundle there are two conduction electrons per unit cell, leads to the electron density $\sim$ 1.7$\times$10$^{3}$ $e/\AA^3$. It is substantial smaller than 8.6$\times$10$^{3}$ $e/\AA^3$ in Cs solid. 
\bibitem{20} DMOL is a density functional theory (DFT) package based atomic basis distributed by Accelrys (B.Delley, J.Chem.Phys.{\bf 92}, 508(1990)). 
\bibitem{21} J.P.Perdew and Y.Wang, Phys.Rev.B{\bf 45}, 13244(1992).
\bibitem{22} Y.Wang and J.P.Perdew, Phys.Rev.B{\bf 43}, 8911(1991); {\bf 44}, 13298(1991).
\bibitem{23} H.J.Monkhorst and J.D.Pack, Phys.Rev.B{\bf 13}, 5188(1976).
\bibitem{24} H.B.Michaelson, J.Appl.Phys.{\bf 48}, 4729(1977).
\bibitem{25} J.A.Zimmerman, J.R.Eyler, S.B.H.Bach, S.W.McElvany, J.Chem.Phys.{\bf 94}, 3556(1991).

\bibitem{26} R.Saito, G.Dressehaus, M.S.Dresselhaus, {\em Physics Properties of Carbon Nanotubes}, (World Scientific, New York, 1998).

\bibitem{27} L.C.Venema, J.W.G.Wildoer, J.W.Janssen, S.J.Tans, H.L.J.T.Tuinstra, L.P.Kouwenhoven, C.Dekker, Science{\bf 283}, 52(1999); A.Buldum, J.P.Lu, Phys.Rev.B{\bf 63}, 161403(2001).

\bibitem{28} J.Zhao, A.Buldum, J.Han, J.P.Lu, Phys.Rev.Lett.{\bf 85}, 1706(2000). 

\bibitem{29} P.Delaney, H.J.Choi, J.Ihm, S.G.Louie, M.L.Cohen, Phys.Rev.B{\bf 60}, 7899(1999).

\bibitem{30} J.Kong, C.Zhou, E.Yenilmez, H.Dia, Appl.Phys.Lett.{\bf 77}, 3977(2000); V.Derycke, R.Martel, J.Appenzeller, Ph.Avouris, Nano.Lett.{\bf 1}, 454(2001).

\bibitem{31} M.S.Dresselhaus, G.Dresselhaus, Adv.Phys.{\bf 30}, 139(1981).

\bibitem{32} C.Bower, S.Suzuki, K.Tanigaki, O.Zhou, Appl.Phys.A{\bf 67}, 47(1998); S.Suzuki, C.Bower, O.Zhou, Chem.Phys.Lett.{\bf 285}, 230(1998).
\bibitem{33} C.Bower, A.Kleinhammes, Y.Wu, O.Zhou, Chem.Phys.Lett.{\bf 288}, 481(1998).

\bibitem{34} R.S.Lee, H.J.Kim, J.E.Fischer, A.Thess, R.E.Smalley, Nature {\bf 388}, 255(1997).
\bibitem{35} R.S.Lee, H.J.Kim, J.E.Fischer, J.Lefebvre, M.Radosavljevic, J.Hone, A.T.Johnson, Phys.Rev.B{\bf 61}, 4526(2000).

\bibitem{36} A.M.Rao, P.C.Eklund, S.Bandow, A.Thess, and R.E.Smalley, Nature {\bf 388}, 257(1997).

\bibitem{37} R.Jacquemin, S.Kazaoui, D.Yu, A.Hassanien, N.Minami, H.Kataura, Y.Achiba, Syntheic Metal {\bf 115}, 283(2000); N.Minami, S.Kazaoui, R.Jacquemin, H.Yamawaki, K.Aoki, H.Kataura, Y.Achiba, Syntheic Metal {\bf 116}, 405(2001).
\bibitem{38} W.Yue, private communication

\end{references}
\end{document}